\documentclass[fleqn,10pt]{wlscirep}

\title{Ge$V_n$ complexes for silicon-based room-temperature single-atom nanoelectronics}

\author[1,2,*]{Simona Achilli}
\author[1,2]{Nicola Manini}
\author[1,2]{Giovanni Onida}
\author[3]{Takahiro Shinada}
\author[4]{Takashi Tanii}
\author[5]{Enrico Prati}

\affil[1]{Dipartimento di Fisica, Universit\`a degli Studi di Milano, Via Celoria 16, 20133 Milano, Italy.}
\affil[2] {European Theoretical Spectroscopy Facility (ETSF)}
\affil[*] {Correspondence to:~simona.achilli@unimi.it}
\affil[3]{Center for Innovative Integrated Electronic Systems, Tohoku University, 468-1 Aramaki Aza Aoba, Aoba-ku, 980-8572 Sendai, Japan.}
\affil[4]{Faculty of Science and Engineering, Waseda University, 3-4-1 Ohkubo, Shinjuku, 169-8555 Tokyo, Japan.}
\affil[5]{Istituto di Fotonica e Nanotecnologie, Consiglio Nazionale delle Ricerche, Piazza Leonardo da Vinci 32, 20133 Milano, Italy.}

\begin{abstract}  

We propose germanium-vacancy complexes (Ge$V_n$) as a viable ingredient to exploit single-atom quantum effects in silicon devices at room temperature.
Our predictions, motivated by the high controllability of the location of the defect via accurate single-atom implantation techniques,
are based on ab-initio Density Functional Theory calculations
within a parameter-free screened-dependent hybrid functional scheme,
suitable to provide reliable bandstructure energies
and defect-state wavefunctions.
The resulting defect-related excited states, at
variance with those arising from conventional dopants such as phosphorous,
turn out to be deep enough to ensure device operation
up to room temperature and 
exhibit a far more localized wavefunction.
\end{abstract}
\begin{document}

\flushbottom
\maketitle
%
%
\thispagestyle{empty}

\section*{Introduction}

The developement of on-demand individual deep impurities in silicon is motivated by their employment as a physical substrate for qubits \cite{pla2012single}, for emitting individual photons \cite{aharonovich2016solid}, to fabricate Hubbard-like quantum systems \cite{fratino2017signatures,baczewski2018multiscale}, and to engineer properties of nanometric-scale transistors \cite{shinada2014opportunity}. Electrically-controlled spin qubits in silicon have been reported so far at cryogenic temperature \cite{pla2012single, maurand2016cmos}, while optical control of silicon qubit is still lacking \cite{awschalom2013quantum}.
Highly-correlated electron states in defects such as NV centers in diamond \cite{dolde2013room,NVreview} and divacancies in SiC \cite{koehl2011room,christle2015isolated} can be exploited as room-temperature optically-controlled qubits, thanks to a deep donor state optically coupled to excited states in the band gap. 
In silicon, the di-vacancy structure would be potentially interesting for engineering a similar spectrum, but creating such defect type
on demand in the bulk is currently unfeasible.
Conversely, exploiting Ge
atom implantation in silicon would offer the opportunity of 
correlated and
controlled spatial positioning, thanks to 
the tendency of Ge to 
recombine with 
vacancies.

Single-atom devices based on conventional doping elements such as phosphorous \cite{hamid2013electron, tan2009transport}, arsenic \cite{prati2014atomic, hori2012quantum} and boron \cite{khalafalla2007identification}, as well as other shallow-level dopants \cite{schenkel2006electrical, van2015single} are limited by their shallow impurity electronic ground state
($\sim 40-50$~meV from the conduction or the valence band edge), so they become fully ionized as soon as one raises the temperature above $\sim 15-20$~K.
%
%
Room-temperature transport across a disordered 1-dimensional array of P donors implanted in a silicon transistor channel has been demonstrated\cite{prati2016band}.
However, in order to secure bound electrons to an isolated donor at room temperature or to electrically manipulate spin states up to $5-10$~K,
it is crucial to rely on deep impurity states near the middle of the band gap.
Deep levels in the silicon bandgap can be generated by electron irradiation of silicon doped by As, P, and Sb atoms \cite{Watk64, Lars06} but the lack of position control and their low annealing temperature between 350 and 450~K make them unsuitable for microelectronic processes.

Isovalent impurities in silicon for accessing the high-temperature regime have been explored\cite{mori2014band,mori2015study}.
Germanium, when dissolved in a substitutional position, does not generate any useful localized state, being isovalent to silicon. A careful choice of the annealing temperature after
implantation around 750~K, \cite{Supr95} however, allows one to activate the defect forming deep energy states in the silicon band gap, associated to germanium-vacancy complexes (Ge$V_n$).

The localized levels of these Ge$V_n$ defects \cite{Supr95,Shul73} have been characterized in the 1970's by deep level transient spectroscopy (DLTS) showing two energy states around $-0.53$~eV and $-0.28$~eV below the conduction-band minimum.
These energy levels are similar to those reported
for the 
simple
silicon vacancy, that would be suitable to behave as deep donor state in terms of energy.
Nevertheless the vacancy location
is not controllable in the
process of device fabrication.
On the contrary, Ge ions can be implanted by single-ion
implantation technique with nm-scale precision. 

The diffusion coefficient of Ge in silicon
is similar to that of Si in silicon,
namely
much lower than, e.g, that of P and As and other deep-level transition metal dopants (Au, Fe, Cu, Ni) \cite{Mehrer},
in particular at the low annealing temperature of 750~K.
The formation and activation of the Ge$V$
complexes is therefore ascribed to the mobility
of the vacancies, which recombine with the Ge ions.
The Ge atom therefore provides the spatial
control by pinning the position of the vacancy, which, in turn, provides the energy
level deep in the bandgap.
As the formation yield may be as low  as around 10\%,
similarly to the case of N$V$-centers and Si$V$
in diamond, the implantation of 
a countable number of atoms by
single-ion implantation may achieve the desired
number of Ge$V$ defects per implantation site.
Quantum devices based on such properties may range 
from room-temperature 2D Hubbard 
systems to single-defect-based transistors.

Our work has been triggered by the availability of single-atom implantation techniques as developed by two of us.
Such techniques have already been used to deal with ions such as P\cite{jamieson2005controlled},
As\cite{prati2012anderson}, Bi\cite{weis2012electrical}, C\cite{tamura2014array},
Ge\cite{prati2015single}, and Er\cite{celebrano20171}, by implanting them one-by-one in a controlled way\cite{shinada2016deterministic}.
Because of the wide availability of Ge in
microelectronics processes 
and its role in controlling the position of deep-level defects 
\cite{prati2015single}, Ge represents a promising candidate
for the extension of single-atom technology to high temperature, with the advantage of a relatively straightforward integration with the standard fabrication technologies of conventional microelectronics.

We characterize here different Ge$V_n$ complexes in silicon, namely a substitutional Ge atom bound to one, two or three adjacent vacancies, and analyze, from a theoretical perspective, their local arrangement and electronic properties.
Previous theoretical works have already proven the tendency of Ge to cluster with vacancies to relax strain, and the resulting stability of Ge$V_n$ complexes\cite{Chen08,Chro09,Vanh10}.

The calculation of the electronic properties of the systems considered here cannot
exploit simple one-electron theories, such as the effective-mass theory used for
conventional donors, and requires
higher-level {\it ab-initio} calculations,
usually performed in the Density Functional (DFT) approach \cite{Overhof}.

In this context,
the theoretical description of shallow donors has to
manage the issue of the large extension of the defect states 
that cannot be correctly described in
manageably small computational cells.
Differently, deep energy levels are expected to be localized around the defect with wavefunctions that decay in a range of a few atomic units.
In the latter cases
spurious delocalization effects in the theoretical description
can arise due to the approximate local or semi-local exchange
and correlation potential usually adopted in DFT.
These unphysical effects are 
a consequence of the self-interaction error which can be corrected
by many-body treatments or high-level functionals.
These methods correct at the same time the localization of the wavefunctions and 
the binding energies of the localized levels, by solving the 
gap problem encountered in DFT that would require otherwise alternative
procedures to estimate the binding energies of defect states, as recently
shown for example by J.\ S.\ Smith {\it et al.}\cite{Smit17}
Specifically, here
we adopt a DFT approach based on a screening-dependent non-local exchange term
that corrects the self-interaction error inherent in local/semilocal approximations,
providing a reliable estimate of the silicon gap and of the energy position
of the defect states relative to the conduction-band minimum\cite{Sko14}.
This kind of hybrid functional exploits an analytical expression
for the exact exchange fraction, which is inversely proportional
to the macroscopic electronic dielectric constant of the material.
In this way the effective screening of the long-range tail of the
Coulomb potential is naturally accounted for in the {\it ab initio}
procedure, leading to excellent results
in reproducing the electronic
properties of nonmetallic systems\cite{Sko14,Gero15,Gero15b}.

With the experimental band-gap value excellently reproduced by the
adopted hybrid functional, the excited states in the gap, obtained here
in terms of charge transition levels (CTLs), are derived directly
from the eigenvalues of the neutral and excited system in the spirit
of Janak's theorem\cite{Jana78}.

\section*{Results and Discussion}

\begin{figure}
\includegraphics[width=1.0\columnwidth]{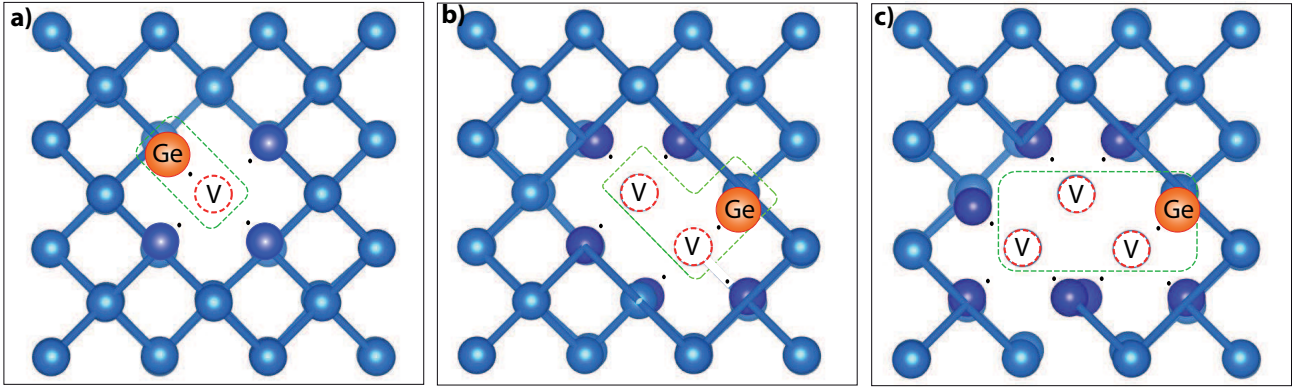}
\caption{The relaxed geometries of the three analyzed complexes: Ge$V$, Ge$V_2$, Ge$V_3$. 
Silicon atoms surrounding directly the vacancy are highlighted by a darker color.
The dots near these atoms indicate the unpaired electrons in the dangling bonds.
}
 \label{geometry}
\end{figure}

Figure~\ref{geometry} shows the relaxed atomic configuration around the defect
complexes considered here.
Relaxation from the ideal crystal geometry leads to quite small displacements, mainly involving atoms surrounding the vacancy which move towards the void. The largest displacement is shown by the Ge atom, as expected.
The tendency of Ge to aggregate to vacancy complexes is confirmed by the defect binding energies, reported in the Supplementary Information, which agree well with previous calculations\cite{Chro09}.

Focusing on the electronic properties, a substitutional Ge atom does not introduce any doping charge in the crystal: the number of unpaired electrons (black small dots in Fig.~\ref{geometry}) associated to any Ge$V_n$ defect complex is even, being the same of the corresponding $V_n$ cluster in silicon.
According to our simulations, the most stable spin configuration is a global $S=0$ state, although with a nonzero spin density appearing locally on the atoms surrounding the defect  and rapidly decaying away from it.

To identify the dominant type of defects one cannot trust a purely equilibrium stability analysis based on the binding energy and on mass-action analysis \cite{Kro56}, because of the complex kinetic effects that direct the Ge$V$ aggregation process, and the strong damage effects due to the Ge-implantation technique.
A theoretical prediction of the relative abundance of the different Ge$V_n$ defects would require to account for all such effects, and is beyond the scope of the present work.
On the other hand, the single-vacancy complex Ge$V$ is certainly stable, and has been identified in experimental studies as the likely source of the observed DLTS signal\cite{Supr95,Shul73}.
We hence postulate that the dominant defect type is indeed the Ge$V$ complex.

\begin{figure} 
\includegraphics[width=0.8\columnwidth]{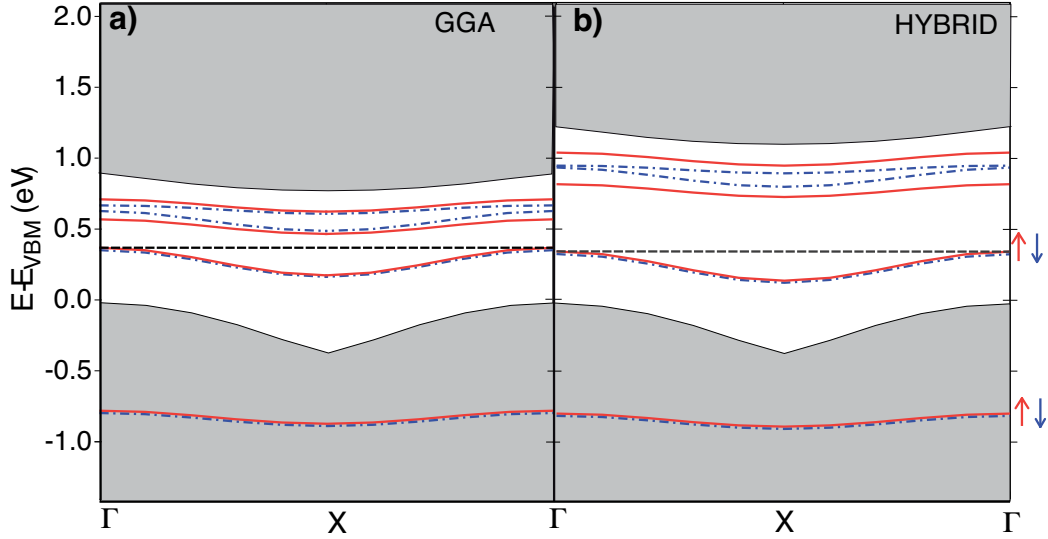}
\caption{
DFT electronic structure of the Ge$V$ defect obtained with different approximations for the exchange-correlation energy: (a) GGA (PBE) and (b) a hybrid functional.
Red (solid) and blue (dot-dashed) lines and arrows identify the localized defect levels for the majority and minority spin components, respectively.
The dashed line marks the Fermi level.
Gray areas correspond to the bulk silicon bands.
The residual $k$-dispersion of the impurity states is an artifact of the relatively small size of the supercell.}
  \label{bands}
\end{figure}

In Ge$V$ two electrons from the dangling bonds
settle in a deep level within the valence band,
similar to the 
$a_1$
(s-type) state of the bare Si vacancy\cite{Wat86}.
Further dangling-bond electrons progressively fill
higher-energy defect states, which appear inside the band gap. 
Figure~\ref{bands}b reports the computed supercell band-structure along the $\Gamma$-X direction.
Majority and minority spin defect states are shown as red (solid) and blue (dot-dashed) lines. For comparison, we also show in Fig.~\ref{bands}a the results obtained by using the standard GGA functional\cite{PBE}.

The system hence exhibits eight localized states (four for each spin projection).
Globally, these states host the four electrons from the dangling bonds,
so that within the band gap only the two lowest defect states are occupied,
while the higher localized levels sit above the Fermi energy
(dashed horizontal line) and are hence empty in the ground-state configuration.
Importantly, the comparison between Fig.~\ref{bands}a and
Fig.~\ref{bands}b shows that using the hybrid functional,
besides solving the well-known gap-underestimation problem,
it corrects the energy position of the defect states relative
to the valence-band edge in a way that is not reproducible by a simple scissor operator.

As the single-electron Kohn-Sham levels are not representative
of the excited states related to the actual addition or removal
of electrons, we need to compute the appropriate 
charge transition levels \cite{Frey14,Vand04} instead:
\begin{equation} \label{ctl:eq}
\epsilon(q|q')=\frac{E_{{\rm f},q'}-E_{{\rm f},q}}{(q-q')}-E_{\rm CBm}
\,.
\end{equation}
Here $E_{{\rm f},q_i}$ is the formation energy
of the defect in the charge state $q_i$ and
$E_{\rm CBm}$ the conduction band minimum.
The expression \eqref{ctl:eq} includes the total energy
of the involved charged defects, and would therefore
require a correction to eliminate the spurious electrostatic
interaction between the periodic replica of the charged defect, as was proposed in the literature\cite{Lany08,Mako95,Castle06}.
Moreover this total energy may be ill-defined because
of the interaction with the balancing background of charge
which is introduced in the present computational approach
to preserve the global system neutrality.
In order to overcome such an issue we use
Janak's theorem \cite{Jana78} in the Slater approximation,
which allows one to estimate the excitation energy
due to the addition/removal of electrons to/from a defect
state as the mean value of the eigenvalue relative to the
first unoccupied/last occupied energy level before and after the excitation:
\begin{equation}\label{janak:eq}
\frac{E_{{\rm f},q'}-E_{{\rm f},q}}{(q-q')}
=
\frac{(\epsilon_{N+1}-\epsilon_{N})}{2(q-q')}
\end{equation}
where the eigenvalue are referred to the conduction band.

Here we consider the ``thermodynamic'' charge transition
levels, i.e., we account for the geometrical relaxation
of the system besides the electronic one, at the new charge state.
This choice is motivated by the long timescale of electron motions in these systems, with respect to the typical structural relaxation timescale.
For an overview of the atomic displacements induced by charge transitions,
and for the value of the ``adiabatic'' charge transition levels
(i.e., computed without structural relaxation in the charged state),
see the Supplementary Information.

\begin{figure}
\includegraphics[width=0.8\columnwidth]{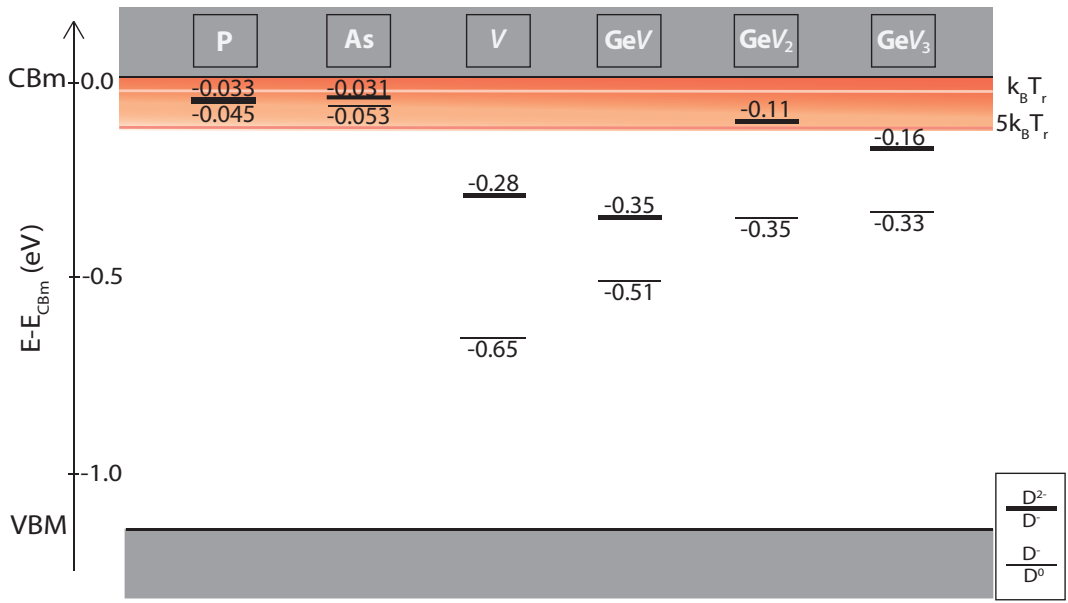}
\caption{ 
\label{CTL}
Charge transition levels of Ge$V_n$ complexes  compared to those of conventional dopant atoms and those of a bare silicon vacancy \cite{Luk03}.
The reference for the energy scale is the conduction-band minimum (CBm).
Red horizontal line: the thermal energy at room temperature ($k_BT_r$).
The shaded area highlights the thermal-excitation probability in a $5k_BT_r$-wide region below the CBm.
Charge transition levels corresponding to the excitation of one and two electron are identified by thin and thick lines, respectively. The charge transition levels of P and As are taken from Ref. \cite{Jaganath} while those of the single vacancy from Ref. \cite{Luk03}}
\end{figure}

Figure~\ref{CTL} reports the computed charge transition
levels for the three defect states analyzed here,
in comparison with the literature values for
conventional dopants and the single vacancy\cite{Jaga81,Luk03,Agga65}.
The excited states corresponding to the D$^{0}$/D$^{-}$ and D$^{-}$/D$^{2-}$ charge transitions of the defect are marked by thin and thick lines, respectively.

While P and As ions are known to produce shallow levels,
whose energy is so close to the conduction band that
these defects are fully ionized at room temperature
(the $k_BT_r$ energy is marked by a red line in Fig.~\ref{CTL}),
a single vacancy in silicon gives rise to deep levels
that would allow single-electron transport at high temperature. 
Unfortunately vacancies in silicon are hard to handle
and control from an experimental point of view.
We find that Ge$V_n$ complexes, easier to control
experimentally, display excited states deep in energy.
In particular those of Ge$V$ are similar to those of the single vacancy. 

For instance, similarly to previous reports on P donor at cryogenic temperature, one could control the charge state of one Ge$V$ defect at room temperature which in turn electrostatically controls a nanometric size room temperature single electron transistor. This can be done by placing the donor defect in the proximity of the channel and by controlling it by means of a side gate \cite{mazzeo2012charge} or by photonic processes \cite{moraru2011atom}.

Our calculated thermodynamic charge transition levels for this defect,
equal to $-0.51$ and $-0.35$~eV for the transition
from D$^{0}$ to D$^{-}$ and from D$^{-}$ to D$^{2-}$,
are in good agreement with the measured DLTS
levels observed after Ge implantation at $-0.53$ eV and $-0.28$ eV respectively,
suggesting that this defect is
indeed
present in the Ge implanted sample\cite{Supr95}.
The deviation from experimental observations by DLTS
at $-0.28$~eV \cite{Supr95,Shul73,Budt98} could be
explained by a partial occupation and a consequent
transient non-stationary condition
(see Fig. S.2 in the Supplementary Information).
For the sake of completeness, the experimental value
is also compatible with the Ge$V_2$ D$^0/$D$^-$ transition.
Indeed, as indicated in Fig.~\ref{CTL}, larger Gs$V_{n\geq 2}$
defects exhibit shallower levels, with only the D$^{0}/$D$^{-}$
charge excitation sufficiently deep to ensure
the trapping of electrons at room temperature.

\begin{figure}
\includegraphics[width=0.8\columnwidth]{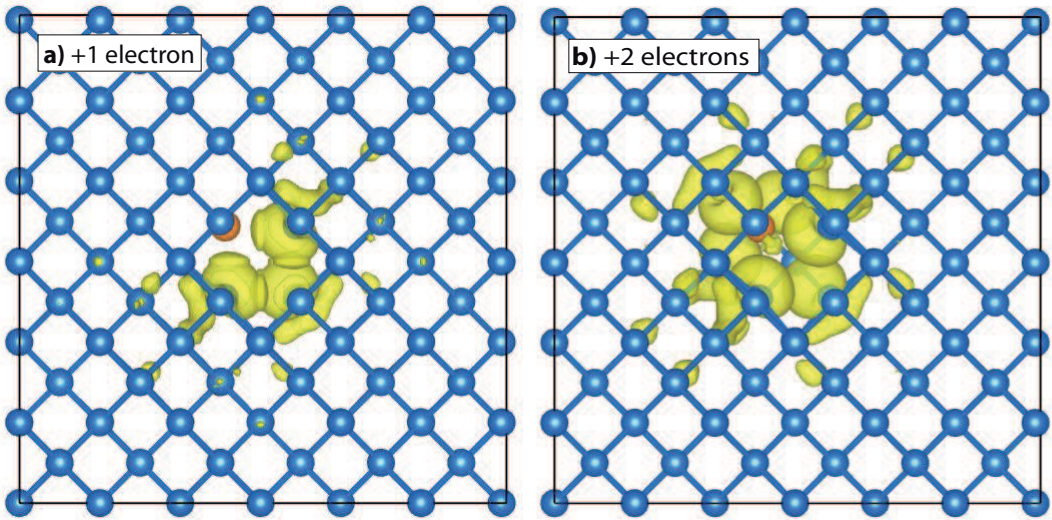}
\caption{\label{WF}
Iso-surfaces (yellow, iso-level $= 0.001$ electrons/\AA$^3$)
for the electronic density of the two lowest
negative ionization states of Ge$V$, namely
(a) at $-0.51$~eV and (b) at $-0.35$~eV, corresponding to D$^-$ and D$^{2-}$, respectively. 
The density for 2 bound electrons differs from that of 1 electron by far more than a pure factor 2: it takes full (electronic and structural) relaxation into account, in particular as induced by the electron-electron Coulomb repulsion. 
The side of the simulation cell (black square) is 1.64 nm.
}
\end{figure}

\begin{figure}
\includegraphics[width=0.7\columnwidth]{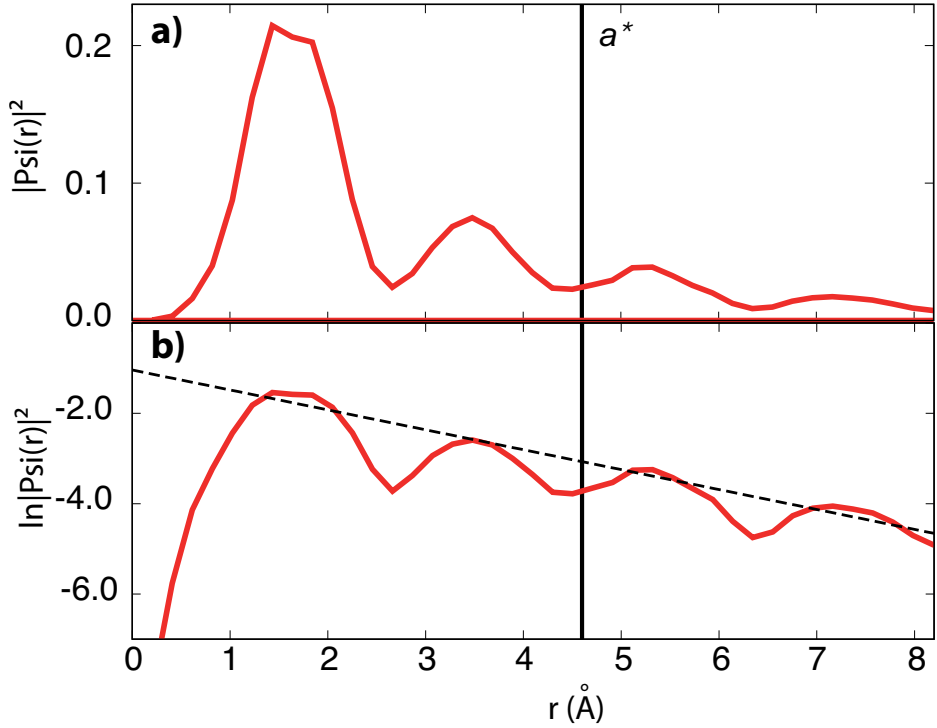}
\caption{  \label{Bohr}
(a) Spherically-averaged charge density for the D$^{-}$ charge state of Ge$V$ as a function of the radial distance from the defect.
(b) The natural logarithm of the same quantity, compared to
the best fit with an exponential as in Eq.~\eqref{fit:eq} (dashed line).
The decay length $a^*$ is marked by a vertical line.
}
\end{figure}

Figure~\ref{WF} displays the charge density of the first
unoccupied electronic state in the gap,
when filled by one (a) and two (b) electrons, respectively.
Notably, Fig.~\ref{WF}a corresponds also to the spin density of the D$^{-}$ charge state.
Differently, the spin density for the closed-shell D$^{2-}$ charge state is zero.
This picture evidences the localization of the additional
electrons on the defect, being the charge density substantially
decayed outside a radius of 0.5~nm away from the defect.
This is confirmed also by the radial decay of the excited-state wavefunction, whose 
spherical average we report in Figure~\ref{Bohr}.
This figure also shows a fit of the envelop
of the wavefunction with an exponential function
\begin{equation} \label{fit:eq}
|\psi(r)|^2 \simeq A*\exp(-2 r/a^*)
\,.
\end{equation}
For the Ge$V$ complex we obtain an effective decay length
$a^* \simeq 0.46$~nm, a value much
smaller than
for the shallow states of conventional dopant atoms.
This greater localization
is expected for deep energy levels, in which electrons
are retained much closer to the defect center.

In summary, a Ge$V_n$ defect in silicon behaves as an isovalent donor atom, carrying a deep empty state in the silicon band gap.
For the purpose of its employment for room-temperature single-atom nanoeletronics,
it combines the deep levels of the $V$ vacancy with the spatial control possible by single-ion 
implantation,
thanks to the tendency of Ge and the vacancy $V$ to form a complex
at an annealing temperature around 750~K.

Because of the relatively low annealing temperature required to activate the Ge$V_n$ complexes, such process step would be necessarily performed after standard annealing of standard diffusion of contacts and charge reservoirs which involves high temperature.

The adoption of screened exchange hybrid functionals was crucial for simulating
Ge$V_n$ complexes in silicon, as this method
allowed us to determine reliably not only the local
geometry of such defects, but also their electronic properties.
While the ground state lies in the valence band, the
relevant donor state of the Ge$V$ has energy $-0.51$~eV,
in good accord with experiments.
At such energy the defect has a transition from
neutral D$^0$ to negative D$^-$ charge state.
We calculate the charge transition level of the D$^-$/D$^{2-}$
states at $-0.35$~eV in the thermodynamic limit. 
By comparison, our calculations for the 
charge transition levels of the
Ge$V_2$ and the Ge$V_3$
complexes yield corresponding energies which are 
smaller
by a factor $\sim 2$ (see Figure~\ref{CTL}).

The decay length $a^*_0=0.45$~nm of the Ge$V^-$
wavefunction indicates that a significant
overlap of such defect with other similar defects or
with the contacting electrodes in a device require nm-scale
spacing, which is now accessible by the current
semiconductor technology node at 7~nm and below.
The main conclusion of the present work is that Ge$V$
is a valid candidate to achieve single-atom nanoelectronics
at noncryogenic temperature, thanks to its deep excited
state in the band gap, which can keep an electron trapped even at room temperature.

\section*{Methods}

We carried out the first-principles calculations in an
all-electron DFT formalism based on linear combination
of atomic orbitals and a Gaussian-type basis set,
as implemented in the CRYSTAL14 code \cite{CRY14}.
The electronic exchange and correlation was included via a hybrid functional:
\begin{equation}
\label{hyb}
v_{\rm XC}(\mathbf r, \mathbf {r'}) = \alpha v_{\rm X}^{\rm EX}(\mathbf{r},\mathbf{r'})+(1-\alpha)v_{\rm X}^{\rm GGA}(\mathbf{r})+v_{\rm C}^{\rm GGA}(\mathbf{r}),
\end{equation}
in which the fraction $\alpha$ of non-local exchange 
is given by the inverse of the static dielectric constant\cite{Alka11,Sko14}.
Such relation can be proved in the framework of the many-body
perturbation theory, considering the Coulomb-hole-plus-screened-exchange
approximation (COHSEX) for the GW self energy \cite{Hedi65}
in the static limit ($\omega=0$):
\begin{equation}
\Sigma_{\rm GW}(\mathbf r, \mathbf {r'},0) =
\Sigma_{\rm COH}(\mathbf{r},\mathbf{r'})+\Sigma_{\rm SEX}(\mathbf{r},\mathbf{r'}),
\end{equation}
where the local $\Sigma_{\rm COH}(\bf{r},\bf{r'})$ term accounts
for the interaction between the electron and the static
polarization cloud.
The non-local $\Sigma_{\rm SEX}(\bf{r},\bf{r'})$
is the static screened exchange:
\begin{eqnarray}
\label{COH}
\Sigma_{\rm COH}(\mathbf{r},\mathbf{r'})&=&
-\frac 12 \delta(\mathbf{r}-\mathbf{r'})[v(\mathbf{r},\mathbf{r'})-W(\mathbf{r},\mathbf{r'})],
\\
\label{SEX}
\Sigma_{\rm SEX}(\mathbf{r},\mathbf{r'})&=&
-\sum\limits_{i=1}^{N_{\rm occ}}\phi(\mathbf{r})\phi^*(\mathbf{r'})W(\mathbf{r},\mathbf{r'}).
\end{eqnarray}

The screened Coulomb potential
\begin{equation}
\label{W}
W(\mathbf{r},\mathbf{r'}) = \int \mathbf{dr''} \frac{v(\mathbf{r''},\mathbf{r'})}{\epsilon(\mathbf{r},\mathbf{r'})}
\end{equation}
in Eq.~\eqref{COH} and \eqref{SEX}
can be evaluated by neglecting the microscopic
component of the dielectric screening and considering
the macroscopic dielectric function
$\epsilon_\infty$ instead of the microscopic one.

With this choice the COH and SEX contributions
to the self energy correspond to the local and
non-local exchange contributions in equation
\eqref{hyb} and $\alpha=1/\epsilon_\infty$.

Such an expression for $\alpha$
was proved to be suitable to reproduce the electronic gaps
of oxides and semiconductors, and the excited states
therein, with uncommon accuracy\cite{Gero15,Gero15b}.
Moreover, by limiting the unphysical wavefunction
delocalization mainly attributable to self-interaction effects
plaguing local exchange-correlation potential approximations, 
the screened exchange hybrid potential allows us
to describe the spatial decay of the localized defect
wavefunctions accurately.
Within the present scheme, one gains the additional
practical advantage of a reduced need for huge simulation
supercells, usually adopted in order to limit the
mutual interaction among periodic defect replicas\cite{Smit17}.

We considered here one defect in a $3 \times 3 \times 3$
simple cubic silicon supercell, i.e., a cubic cell
with a theoretical
lattice constant $a=5.46$~\AA~containing
216 Si atoms in absence of vacancies.
All Ge$V_n$ systems have been structurally relaxed,
until the maximum and the root mean square of the
residual forces reduced to 1.4 and 0.9~pN, respectively.
We used a Monkhorst-Pack grid of $4\times4\times4$
$k$-points and the basis set
for Si and Ge proposed by Towler\cite{Towler}.

\section*{Acknowledgements}

The authors acknowledge financial support of the NFFA infrastructure under Project ID 188,
the CINECA
supercomputing grant
project HPL13P8AUS
under the LISA initiative,
and R. Dovesi, S. Casassa and S. Salustro
for software support.
The authors acknowledge also the CNR-JSPS Bilateral Seminar Grant of 2017 and Grant-in-Aid for Scientific Research (nos. 16K14242 and 18H03766) from MEXT, Japan.
E. P. acknowledges the PEST 2010-2012 Ministero Affari Esteri (MAE), Italy, the Short Term Mobility Program of 2013 of CNR, and the JSPS Invitation Fellowship Program 2014 at Waseda University.
G.O. also acknowledges Matteo Gerosa for useful discussions.

\section*{Author contributions statement}
E. P., T. S and T. T. defined the design of the structure according to experimental constraints. E. P., G. O. and T. T. coordinated the research.
S.A. performed first-principles DFT calculations and defined the theoretical setup and approach with N.M..
All the authors contributed to discuss the results and to prepare the manuscript.

\section*{Additional information}
\textbf{Competing financial interests:} The authors declare that they have no competing interests.
The corresponding author is responsible for submitting a \href{http://www.nature.com/srep/policies/index.html#competing}{competing financial interests statement} on behalf of all authors of the paper. 

\noindent\textbf{Data availability statement:}The datasets generated during and/or analysed during the current study are available from the corresponding author on reasonable request.


\begin{thebibliography}{10}
\expandafter\ifx\csname url\endcsname\relax
  \def\url#1{\texttt{#1}}\fi
\expandafter\ifx\csname urlprefix\endcsname\relax\def\urlprefix{URL }\fi
\expandafter\ifx\csname doiprefix\endcsname\relax\def\doiprefix{DOI }\fi
\providecommand{\bibinfo}[2]{#2}
\providecommand{\eprint}[2][]{\url{#2}}

\bibitem{pla2012single}
\bibinfo{author}{Pla, J.~J.} \emph{et~al.}
\newblock \bibinfo{journal}{\bibinfo{title}{A single-atom electron spin qubit
  in silicon}}.
\newblock {\emph{\JournalTitle{Nature}}} \textbf{\bibinfo{volume}{489}},
  \bibinfo{pages}{541} (\bibinfo{year}{2012}).

\bibitem{aharonovich2016solid}
\bibinfo{author}{Aharonovich, I.}, \bibinfo{author}{Englund, D.} \&
  \bibinfo{author}{Toth, M.}
\newblock \bibinfo{journal}{\bibinfo{title}{Solid-state single-photon
  emitters}}.
\newblock {\emph{\JournalTitle{Nature Photonics}}}
  \textbf{\bibinfo{volume}{10}}, \bibinfo{pages}{631} (\bibinfo{year}{2016}).

\bibitem{fratino2017signatures}
\bibinfo{author}{Fratino, L.}, \bibinfo{author}{Semon, P.},
  \bibinfo{author}{Charlebois, M.}, \bibinfo{author}{Sordi, G.} \&
  \bibinfo{author}{Tremblay, A.-M.}
\newblock \bibinfo{journal}{\bibinfo{title}{Signatures of the mott transition
  in the antiferromagnetic state of the two-dimensional hubbard model}}.
\newblock {\emph{\JournalTitle{Physical Review B}}}
  \textbf{\bibinfo{volume}{95}}, \bibinfo{pages}{235109}
  (\bibinfo{year}{2017}).

\bibitem{baczewski2018multiscale}
\bibinfo{author}{Baczewski, A.} \emph{et~al.}
\newblock \bibinfo{journal}{\bibinfo{title}{Multiscale modeling of dopant
  arrays in silicon}}.
\newblock {\emph{\JournalTitle{Bulletin of the American Physical Society}}}
  (\bibinfo{year}{2018}).

\bibitem{shinada2014opportunity}
\bibinfo{author}{Shinada, T.} \emph{et~al.}
\newblock \bibinfo{title}{Opportunity of single atom control for quantum
  processing in silicon and diamond}.
\newblock In \emph{\bibinfo{booktitle}{Silicon Nanoelectronics Workshop
  (SNW)}}, \bibinfo{pages}{1--2} (\bibinfo{organization}{IEEE},
  \bibinfo{year}{2014}).

\bibitem{maurand2016cmos}
\bibinfo{author}{Maurand, R.} \emph{et~al.}
\newblock \bibinfo{journal}{\bibinfo{title}{A {CMOS} silicon spin qubit}}.
\newblock {\emph{\JournalTitle{Nature communications}}}
  \textbf{\bibinfo{volume}{7}}, \bibinfo{pages}{13575} (\bibinfo{year}{2016}).

\bibitem{awschalom2013quantum}
\bibinfo{author}{Awschalom, D.~D.}, \bibinfo{author}{Bassett, L.~C.},
  \bibinfo{author}{Dzurak, A.~S.}, \bibinfo{author}{Hu, E.~L.} \&
  \bibinfo{author}{Petta, J.~R.}
\newblock \bibinfo{journal}{\bibinfo{title}{Quantum spintronics: engineering
  and manipulating atom-like spins in semiconductors}}.
\newblock {\emph{\JournalTitle{Science}}} \textbf{\bibinfo{volume}{339}},
  \bibinfo{pages}{1174--1179} (\bibinfo{year}{2013}).

\bibitem{dolde2013room}
\bibinfo{author}{Dolde, F.} \emph{et~al.}
\newblock \bibinfo{journal}{\bibinfo{title}{Room-temperature entanglement
  between single defect spins in diamond}}.
\newblock {\emph{\JournalTitle{Nature Physics}}} \textbf{\bibinfo{volume}{9}},
  \bibinfo{pages}{139} (\bibinfo{year}{2013}).

\bibitem{NVreview}
\bibinfo{author}{Schirhagl, R.}, \bibinfo{author}{Chang, K.},
  \bibinfo{author}{Loretz, M.} \& \bibinfo{author}{Degen, C.~L.}
\newblock \bibinfo{journal}{\bibinfo{title}{Nitrogen-vacancy centers in
  diamond: Nanoscale sensors for physics and biology}}.
\newblock {\emph{\JournalTitle{Annual Review of Physical Chemistry}}}
  \textbf{\bibinfo{volume}{65}}, \bibinfo{pages}{83--105}
  (\bibinfo{year}{2014}).

\bibitem{koehl2011room}
\bibinfo{author}{Koehl, W.~F.}, \bibinfo{author}{Buckley, B.~B.},
  \bibinfo{author}{Heremans, F.~J.}, \bibinfo{author}{Calusine, G.} \&
  \bibinfo{author}{Awschalom, D.~D.}
\newblock \bibinfo{journal}{\bibinfo{title}{Room temperature coherent control
  of defect spin qubits in silicon carbide}}.
\newblock {\emph{\JournalTitle{Nature}}} \textbf{\bibinfo{volume}{479}},
  \bibinfo{pages}{84} (\bibinfo{year}{2011}).

\bibitem{christle2015isolated}
\bibinfo{author}{Christle, D.~J.} \emph{et~al.}
\newblock \bibinfo{journal}{\bibinfo{title}{Isolated electron spins in silicon
  carbide with millisecond coherence times}}.
\newblock {\emph{\JournalTitle{Nature materials}}}
  \textbf{\bibinfo{volume}{14}}, \bibinfo{pages}{160} (\bibinfo{year}{2015}).

\bibitem{hamid2013electron}
\bibinfo{author}{Hamid, E.} \emph{et~al.}
\newblock \bibinfo{journal}{\bibinfo{title}{Electron-tunneling operation of
  single-donor-atom transistors at elevated temperatures}}.
\newblock {\emph{\JournalTitle{Phys. Rev. B}}} \textbf{\bibinfo{volume}{87}},
  \bibinfo{pages}{085420} (\bibinfo{year}{2013}).

\bibitem{tan2009transport}
\bibinfo{author}{Tan, K.~Y.} \emph{et~al.}
\newblock \bibinfo{journal}{\bibinfo{title}{Transport spectroscopy of single
  phosphorus donors in a silicon nanoscale transistor}}.
\newblock {\emph{\JournalTitle{Nano Lett.}}} \textbf{\bibinfo{volume}{10}},
  \bibinfo{pages}{11--15} (\bibinfo{year}{2009}).

\bibitem{prati2014atomic}
\bibinfo{author}{Prati, E.} \& \bibinfo{author}{Shinada, T.}
\newblock \bibinfo{title}{Atomic scale devices: Advancements and directions}.
\newblock In \emph{\bibinfo{booktitle}{Electron Devices Meeting (IEDM)}},
  \bibinfo{pages}{1--2} (\bibinfo{organization}{IEEE International},
  \bibinfo{year}{2014}).

\bibitem{hori2012quantum}
\bibinfo{author}{Hori, M.}, \bibinfo{author}{Shinada, T.},
  \bibinfo{author}{Guagliardo, F.}, \bibinfo{author}{Ferrari, G.} \&
  \bibinfo{author}{Prati, E.}
\newblock \bibinfo{title}{Quantum transport property in {FET}s with
  deterministically implanted single-arsenic ions using single-ion
  implantation}.
\newblock In \emph{\bibinfo{booktitle}{Silicon Nanoelectronics Workshop
  (SNW)}}, \bibinfo{pages}{1--2} (\bibinfo{organization}{IEEE},
  \bibinfo{year}{2012}).

\bibitem{khalafalla2007identification}
\bibinfo{author}{Khalafalla, M.}, \bibinfo{author}{Ono, Y.},
  \bibinfo{author}{Nishiguchi, K.} \& \bibinfo{author}{Fujiwara, A.}
\newblock \bibinfo{journal}{\bibinfo{title}{Identification of single and
  coupled acceptors in silicon nano-field-effect transistors}}.
\newblock {\emph{\JournalTitle{Appl. Phys. Lett.}}}
  \textbf{\bibinfo{volume}{91}}, \bibinfo{pages}{263513}
  (\bibinfo{year}{2007}).

\bibitem{schenkel2006electrical}
\bibinfo{author}{Schenkel, T.} \emph{et~al.}
\newblock \bibinfo{journal}{\bibinfo{title}{Electrical activation and electron
  spin coherence of ultralow dose antimony implants in silicon}}.
\newblock {\emph{\JournalTitle{Appl. Phys. Lett.}}}
  \textbf{\bibinfo{volume}{88}}, \bibinfo{pages}{112101}
  (\bibinfo{year}{2006}).

\bibitem{van2015single}
\bibinfo{author}{{Van Donkelaar}, J.} \emph{et~al.}
\newblock \bibinfo{journal}{\bibinfo{title}{Single atom devices by ion
  implantation}}.
\newblock {\emph{\JournalTitle{J. Phys.: Condens. Matter}}}
  \textbf{\bibinfo{volume}{27}}, \bibinfo{pages}{154204}
  (\bibinfo{year}{2015}).

\bibitem{prati2016band}
\bibinfo{author}{Prati, E.}, \bibinfo{author}{Kumagai, K.},
  \bibinfo{author}{Hori, M.} \& \bibinfo{author}{Shinada, T.}
\newblock \bibinfo{journal}{\bibinfo{title}{Band transport across a chain of
  dopant sites in silicon over micron distances and high temperatures}}.
\newblock {\emph{\JournalTitle{Sci. Rep.}}} \textbf{\bibinfo{volume}{6}},
  \bibinfo{pages}{19704} (\bibinfo{year}{2016}).

\bibitem{Watk64}
\bibinfo{author}{Watkins, C.~G.} \& \bibinfo{author}{Corbett, J.~V.}
\newblock \bibinfo{journal}{\bibinfo{title}{Defects in irradiated silicon:
  Electron paramagnetic resonance and electron-nuclear double resonance of the
  {S}i-{E} center}}.
\newblock {\emph{\JournalTitle{Phys. Rev.}}} \textbf{\bibinfo{volume}{134}},
  \bibinfo{pages}{A1359} (\bibinfo{year}{1964}).

\bibitem{Lars06}
\bibinfo{author}{{Nylandsted Larsen}, A.} \emph{et~al.}
\newblock \bibinfo{journal}{\bibinfo{title}{{E} center in silicon has a donor
  level in the band gap}}.
\newblock {\emph{\JournalTitle{Phys. Rev. Lett.}}}
  \textbf{\bibinfo{volume}{97}}, \bibinfo{pages}{106402}
  (\bibinfo{year}{2006}).

\bibitem{mori2014band}
\bibinfo{author}{Mori, T.} \emph{et~al.}
\newblock \bibinfo{title}{Band-to-band tunneling current enhancement utilizing
  isoelectronic trap and its application to {TFET}s}.
\newblock In \emph{\bibinfo{booktitle}{VLSI Technology (VLSI-Technology):
  Digest of Technical Papers, Symposium on}}, \bibinfo{pages}{1--2}
  (\bibinfo{organization}{IEEE}, \bibinfo{year}{2014}).

\bibitem{mori2015study}
\bibinfo{author}{Mori, T.} \emph{et~al.}
\newblock \bibinfo{journal}{\bibinfo{title}{Study of tunneling transport in
  {S}i-based tunnel field-effect transistors with {ON} current enhancement
  utilizing isoelectronic trap}}.
\newblock {\emph{\JournalTitle{Appl. Phys. Lett.}}}
  \textbf{\bibinfo{volume}{106}}, \bibinfo{pages}{083501}
  (\bibinfo{year}{2015}).

\bibitem{Supr95}
\bibinfo{author}{Suprun-Belevich, Y.} \& \bibinfo{author}{Palmetshofer, L.}
\newblock \bibinfo{journal}{\bibinfo{title}{Deep defect levels and mechanical
  strain in {G}e$^+-$implanted {S}i}}.
\newblock {\emph{\JournalTitle{Nucl. Instr. Methods Phys. Res. B}}}
  \textbf{\bibinfo{volume}{96}}, \bibinfo{pages}{245--248}
  (\bibinfo{year}{1995}).

\bibitem{Shul73}
\bibinfo{author}{Shulz, M.}
\newblock \bibinfo{journal}{\bibinfo{title}{Deep trap levels of ion-implanted
  germanium in silicon measured by {S}chottky contact techniques}}.
\newblock {\emph{\JournalTitle{Appl. Phys. Lett}}}
  \textbf{\bibinfo{volume}{23}}, \bibinfo{pages}{31} (\bibinfo{year}{1973}).

\bibitem{Mehrer}
\bibinfo{author}{Mehrer, H.}
\newblock \emph{\bibinfo{title}{Diffusion in Solids: Fundamentals, Methods,
  Materials, Diffusion-Controlled Processes}}.
\newblock Springer Series in Solid-State Sciences
  (\bibinfo{publisher}{Springer}, \bibinfo{address}{Berlin, Heidelberg},
  \bibinfo{year}{2007}).

\bibitem{jamieson2005controlled}
\bibinfo{author}{Jamieson, D.~N.} \emph{et~al.}
\newblock \bibinfo{journal}{\bibinfo{title}{Controlled shallow single-ion
  implantation in silicon using an active substrate for sub-20-ke{V} ions}}.
\newblock {\emph{\JournalTitle{Appl. Phys. Lett.}}}
  \textbf{\bibinfo{volume}{86}}, \bibinfo{pages}{202101}
  (\bibinfo{year}{2005}).

\bibitem{prati2012anderson}
\bibinfo{author}{Prati, E.}, \bibinfo{author}{Hori, M.},
  \bibinfo{author}{Guagliardo, F.}, \bibinfo{author}{Ferrari, G.} \&
  \bibinfo{author}{Shinada, T.}
\newblock \bibinfo{journal}{\bibinfo{title}{Anderson-{M}ott transition in
  arrays of a few dopant atoms in a silicon transistor}}.
\newblock {\emph{\JournalTitle{Nature Nanotech.}}}
  \textbf{\bibinfo{volume}{7}}, \bibinfo{pages}{443} (\bibinfo{year}{2012}).

\bibitem{weis2012electrical}
\bibinfo{author}{Weis, C.} \emph{et~al.}
\newblock \bibinfo{journal}{\bibinfo{title}{Electrical activation and electron
  spin resonance measurements of implanted bismuth in isotopically enriched
  silicon-28}}.
\newblock {\emph{\JournalTitle{Appl. Phys. Lett.}}}
  \textbf{\bibinfo{volume}{100}}, \bibinfo{pages}{172104}
  (\bibinfo{year}{2012}).

\bibitem{tamura2014array}
\bibinfo{author}{Tamura, S.} \emph{et~al.}
\newblock \bibinfo{journal}{\bibinfo{title}{Array of bright silicon-vacancy
  centers in diamond fabricated by low-energy focused ion beam implantation}}.
\newblock {\emph{\JournalTitle{Appl. Phys. Expr.}}}
  \textbf{\bibinfo{volume}{7}}, \bibinfo{pages}{115201} (\bibinfo{year}{2014}).

\bibitem{prati2015single}
\bibinfo{author}{Prati, E.} \emph{et~al.}
\newblock \bibinfo{title}{Single ion implantation of {G}e donor impurity in
  silicon transistors}.
\newblock In \emph{\bibinfo{booktitle}{Silicon Nanoelectronics Workshop
  (SNW)}}, \bibinfo{pages}{1--2} (\bibinfo{organization}{IEEE},
  \bibinfo{year}{2015}).

\bibitem{celebrano20171}
\bibinfo{author}{Celebrano, M.} \emph{et~al.}
\newblock \bibinfo{journal}{\bibinfo{title}{1.54 $\mu$m photoluminescence from
  {E}r: {O}x centers at extremely low concentration in silicon at 300 k}}.
\newblock {\emph{\JournalTitle{Optics letters}}} \textbf{\bibinfo{volume}{42}},
  \bibinfo{pages}{3311--3314} (\bibinfo{year}{2017}).

\bibitem{shinada2016deterministic}
\bibinfo{author}{Shinada, T.} \emph{et~al.}
\newblock \bibinfo{title}{Deterministic doping to silicon and diamond materials
  for quantum processing}.
\newblock In \emph{\bibinfo{booktitle}{Nanotechnology (IEEE-NANO), 16th
  International Conference on}}, \bibinfo{pages}{888--890}
  (\bibinfo{organization}{IEEE}, \bibinfo{year}{2016}).

\bibitem{Chen08}
\bibinfo{author}{Chen, J.}, \bibinfo{author}{Wu, T.}, \bibinfo{author}{Ma, X.},
  \bibinfo{author}{Wang, L.} \& \bibinfo{author}{Yang, D.}
\newblock \bibinfo{journal}{\bibinfo{title}{Ge-vacancy pair in {G}e-doped
  {C}zochralski silicon}}.
\newblock {\emph{\JournalTitle{J. Appl. Phys.}}}
  \textbf{\bibinfo{volume}{103}}, \bibinfo{pages}{123519}
  (\bibinfo{year}{2008}).

\bibitem{Chro09}
\bibinfo{author}{Chroneos, A.}, \bibinfo{author}{Grimes, R.~W.} \&
  \bibinfo{author}{Bracht, H.}
\newblock \bibinfo{journal}{\bibinfo{title}{Impact of germanium on vacancy
  clustering in germanium-doped silicon}}.
\newblock {\emph{\JournalTitle{J. Appl. Phys}}} \textbf{\bibinfo{volume}{105}},
  \bibinfo{pages}{016102} (\bibinfo{year}{2009}).

\bibitem{Vanh10}
\bibinfo{author}{Vanhellmont, J.}, \bibinfo{author}{Suezawa, M.} \&
  \bibinfo{author}{Yonenaga, I.}
\newblock \bibinfo{journal}{\bibinfo{title}{On the assumed impact of germanium
  doping on void formation in {C}zochralski-grown silicon}}.
\newblock {\emph{\JournalTitle{J. Appl.Phys.}}} \textbf{\bibinfo{volume}{108}},
  \bibinfo{pages}{016105} (\bibinfo{year}{2010}).

\bibitem{Overhof}
\bibinfo{author}{Overhof, H.} \& \bibinfo{author}{Gerstmann, U.}
\newblock \bibinfo{journal}{\bibinfo{title}{Ab initio calculation of hyperfine
  and superhyperfine interactions for shallow donors in semiconductors}}.
\newblock {\emph{\JournalTitle{Phys. Rev. Lett.}}}
  \textbf{\bibinfo{volume}{92}}, \bibinfo{pages}{087602}
  (\bibinfo{year}{2004}).

\bibitem{Smit17}
\bibinfo{author}{Smith, J.~S.} \emph{et~al.}
\newblock \bibinfo{journal}{\bibinfo{title}{Ab initio calculation of energy
  levels for phosphorus donors in silicon}}.
\newblock {\emph{\JournalTitle{Sci. Rep.}}} \textbf{\bibinfo{volume}{7}},
  \bibinfo{pages}{6010} (\bibinfo{year}{2017}).

\bibitem{Sko14}
\bibinfo{author}{Skone, J.~H.}, \bibinfo{author}{Govoni, M.} \&
  \bibinfo{author}{Galli, G.}
\newblock \bibinfo{journal}{\bibinfo{title}{Self-consistent hybrid functional
  for condensed systems}}.
\newblock {\emph{\JournalTitle{Phys. Rev. B}}} \textbf{\bibinfo{volume}{89}},
  \bibinfo{pages}{195112} (\bibinfo{year}{2014}).

\bibitem{Gero15}
\bibinfo{author}{Gerosa, M.} \emph{et~al.}
\newblock \bibinfo{journal}{\bibinfo{title}{Electronic structure and phase
  stability of oxide semiconductors: Performance of dielectric-dependent hybrid
  functional {DFT}, benchmarked against {GW} band structure calculations and
  experiments}}.
\newblock {\emph{\JournalTitle{Phys. Rev. B}}} \textbf{\bibinfo{volume}{91}},
  \bibinfo{pages}{155201} (\bibinfo{year}{2015}).

\bibitem{Gero15b}
\bibinfo{author}{Gerosa, M.}, \bibinfo{author}{{Di Valentin}, C.},
  \bibinfo{author}{Bottani, C.~E.}, \bibinfo{author}{Onida, G.} \&
  \bibinfo{author}{Pacchioni, G.}
\newblock \bibinfo{journal}{\bibinfo{title}{Communication: Hole localization in
  {A}l-doped quartz {S}i{O}$_2$ within ab initio hybrid-functional {DFT}}}.
\newblock {\emph{\JournalTitle{J.Chem. Phys.}}} \textbf{\bibinfo{volume}{143}},
  \bibinfo{pages}{111103} (\bibinfo{year}{2015}).

\bibitem{Jana78}
\bibinfo{author}{Janak, J.~F.}
\newblock \bibinfo{journal}{\bibinfo{title}{Proof that
  $\frac{\ensuremath{\partial}e}{\ensuremath{\partial}{n}_{i}}=\ensuremath{\epsilon}$
  in density-functional theory}}.
\newblock {\emph{\JournalTitle{Phys. Rev. B}}} \textbf{\bibinfo{volume}{18}},
  \bibinfo{pages}{7165--7168} (\bibinfo{year}{1978}).

\bibitem{Kro56}
\bibinfo{author}{Kr\"oger, F.~A.} \& \bibinfo{author}{Vink, V.~J.}
\newblock \emph{\bibinfo{title}{Solid State Physics}} (\bibinfo{publisher}{F.
  Seitz and D. Turnbull}, \bibinfo{address}{Academic New York},
  \bibinfo{year}{1956}), \bibinfo{edition}{3} edn.

\bibitem{Wat86}
\bibinfo{author}{Watkins, G.~D.}
\newblock \emph{\bibinfo{title}{Deep Centres in Semiconductors}}
  (\bibinfo{publisher}{S. T. Pantelides}, \bibinfo{address}{New York},
  \bibinfo{year}{1986}).

\bibitem{PBE}
\bibinfo{author}{Perdew, J.~P.}, \bibinfo{author}{Burke, K.} \&
  \bibinfo{author}{Ernzerhof, M.}
\newblock \bibinfo{journal}{\bibinfo{title}{Generalized gradient approximation
  made simple}}.
\newblock {\emph{\JournalTitle{Phys. Rev. Lett.}}}
  \textbf{\bibinfo{volume}{77}}, \bibinfo{pages}{3865--3868}
  (\bibinfo{year}{1996}).

\bibitem{Frey14}
\bibinfo{author}{Freysoldt, C.} \emph{et~al.}
\newblock \bibinfo{journal}{\bibinfo{title}{First-principles calculations for
  point defects in solids}}.
\newblock {\emph{\JournalTitle{Rev. Mod. Phys.}}}
  \textbf{\bibinfo{volume}{86}}, \bibinfo{pages}{253--305}
  (\bibinfo{year}{2014}).

\bibitem{Vand04}
\bibinfo{author}{{Van de Walle}, C.~G.} \& \bibinfo{author}{Neugebauer, J.}
\newblock \bibinfo{journal}{\bibinfo{title}{First-principles calculations for
  defects and impurities: Applications to {III}-nitrides}}.
\newblock {\emph{\JournalTitle{J. Appl. Phys.}}} \textbf{\bibinfo{volume}{95}},
  \bibinfo{pages}{3851} (\bibinfo{year}{2004}).

\bibitem{Lany08}
\bibinfo{author}{Lany, S.} \& \bibinfo{author}{Zunger, A.}
\newblock \bibinfo{journal}{\bibinfo{title}{Assessment of correction methods
  for the band-gap problem and for finite-size effects in supercell defect
  calculations: Case studies for {Z}n{O} and {G}a{A}s}}.
\newblock {\emph{\JournalTitle{Phys. Rev. B}}} \textbf{\bibinfo{volume}{78}},
  \bibinfo{pages}{235104} (\bibinfo{year}{2008}).

\bibitem{Mako95}
\bibinfo{author}{Makov, G.} \& \bibinfo{author}{Payne, M.~C.}
\newblock \bibinfo{journal}{\bibinfo{title}{Periodic boundary conditions in ab
  initio calculations}}.
\newblock {\emph{\JournalTitle{Phys. Rev. B}}} \textbf{\bibinfo{volume}{51}},
  \bibinfo{pages}{4014} (\bibinfo{year}{1995}).

\bibitem{Castle06}
\bibinfo{journal}{\bibinfo{author}{Castleton, C. W.~M.},
  \bibinfo{author}{H\"oglund, A.} \& \bibinfo{author}{Mirbt, S.}}
\newblock {\emph{\JournalTitle{Phys. Rev. B}}} \textbf{\bibinfo{volume}{73}},
  \bibinfo{pages}{035215} (\bibinfo{year}{2006}).

\bibitem{Luk03}
\bibinfo{author}{Lukjanitsa, V.~V.}
\newblock \bibinfo{journal}{\bibinfo{title}{Energy levels of vacancies and
  interstitial atoms in the band gap of silicon}}.
\newblock {\emph{\JournalTitle{Semiconductors}}} \textbf{\bibinfo{volume}{37}},
  \bibinfo{pages}{422--431} (\bibinfo{year}{2003}).

\bibitem{Jaganath}
\bibinfo{author}{Jagannath, C.}, \bibinfo{author}{Grabowski, Z.~W.} \&
  \bibinfo{author}{Ramdas, A.~K.}
\newblock \bibinfo{journal}{\bibinfo{title}{Linewidths of the electronic
  excitation spectra of donors in silicon}}.
\newblock {\emph{\JournalTitle{Phys. Rev. B}}} \textbf{\bibinfo{volume}{23}},
  \bibinfo{pages}{2082--2098} (\bibinfo{year}{1981}).

\bibitem{Jaga81}
\bibinfo{author}{Jagannath, C.}, \bibinfo{author}{Grabowski, Z.~W.} \&
  \bibinfo{author}{Ramdas, A.~K.}
\newblock \bibinfo{journal}{\bibinfo{title}{Linewidths of the electronic
  excitation spectra of donors in silicon}}.
\newblock {\emph{\JournalTitle{Phys. Rev. B}}} \textbf{\bibinfo{volume}{23}},
  \bibinfo{pages}{2082} (\bibinfo{year}{1981}).

\bibitem{Agga65}
\bibinfo{author}{Aggarwal, R.~L.} \& \bibinfo{author}{Ramdas, A.~K.}
\newblock \bibinfo{journal}{\bibinfo{title}{Optical determination of the
  symmetry of the ground states of group-{V} donors in silicon}}.
\newblock {\emph{\JournalTitle{Phys. Rev.}}} \textbf{\bibinfo{volume}{140}},
  \bibinfo{pages}{A1246} (\bibinfo{year}{1965}).

\bibitem{mazzeo2012charge}
\bibinfo{author}{Mazzeo, G.} \emph{et~al.}
\newblock \bibinfo{journal}{\bibinfo{title}{Charge dynamics of a single donor
  coupled to a few-electron quantum dot in silicon}}.
\newblock {\emph{\JournalTitle{Applied Physics Letters}}}
  \textbf{\bibinfo{volume}{100}}, \bibinfo{pages}{213107}
  (\bibinfo{year}{2012}).

\bibitem{moraru2011atom}
\bibinfo{author}{Moraru, D.} \emph{et~al.}
\newblock \bibinfo{journal}{\bibinfo{title}{Atom devices based on single
  dopants in silicon nanostructures}}.
\newblock {\emph{\JournalTitle{Nanoscale research letters}}}
  \textbf{\bibinfo{volume}{6}}, \bibinfo{pages}{479} (\bibinfo{year}{2011}).

\bibitem{Budt98}
\bibinfo{author}{Budtz-J\o{}rgensen, C.~V.}, \bibinfo{author}{Kringh\o{}j, P.}
  \& \bibinfo{author}{Larsen, A.~N.}
\newblock \bibinfo{journal}{\bibinfo{title}{Deep-level transient spectroscopy
  of the {G}e-vacancy pair in {G}e-doped $n-$type silicon}}.
\newblock {\emph{\JournalTitle{Phys. Rev. B}}} \textbf{\bibinfo{volume}{58}},
  \bibinfo{pages}{1110} (\bibinfo{year}{1998}).

\bibitem{CRY14}
\bibinfo{author}{Dovesi, R.} \emph{et~al.}
\newblock \bibinfo{journal}{\bibinfo{title}{Crystal14: A program for the ab
  initio investigation of crystalline solids}}.
\newblock {\emph{\JournalTitle{Int. J. Quantum Chem.}}}
  \textbf{\bibinfo{volume}{114}}, \bibinfo{pages}{1287--1317}
  (\bibinfo{year}{2014}).

\bibitem{Alka11}
\bibinfo{author}{Alkauskas, A.}, \bibinfo{author}{Broqvist, P.},  \&
  \bibinfo{author}{Pasquarello, A.}
\newblock \bibinfo{journal}{\bibinfo{title}{Defect levels through hybrid
  density functionals: Insights and applications}}.
\newblock {\emph{\JournalTitle{Phys. Status Solidi B}}}
  \textbf{\bibinfo{volume}{248}}, \bibinfo{pages}{775--789}
  (\bibinfo{year}{2011}).

\bibitem{Hedi65}
\bibinfo{author}{Hedin, L.}
\newblock \bibinfo{journal}{\bibinfo{title}{New method for calculating the
  one-particle {G}reen's function with application to the electron-gas
  problem}}.
\newblock {\emph{\JournalTitle{Phys. Rev.}}} \textbf{\bibinfo{volume}{139}},
  \bibinfo{pages}{A796} (\bibinfo{year}{1965}).

\bibitem{Towler}
\bibinfo{author}{Porter, A.}, \bibinfo{author}{Towler, M.} \&
  \bibinfo{author}{Needs, R.}
\newblock \bibinfo{journal}{\bibinfo{title}{Muonium as a hydrogen analogue in
  silicon and germanium; quantum effects and hyperfine parameters}}.
\newblock {\emph{\JournalTitle{Phys. Rev. B}}} \textbf{\bibinfo{volume}{60}},
  \bibinfo{pages}{13534--13546} (\bibinfo{year}{1999}).

\end{thebibliography}
\end{document}